\documentclass[12pt]{article}
\usepackage{graphicx}

\newcommand\pubnumber{SNSN-XXX-YY}
\newcommand\pubdate{November 15,2014}

\def\napoli{Istituto Nazionale di Fisica Nucleare (INFN), sez. Perugia \\
Via A. Pascoli, I-06123 Perugia, ITALY}
\def\support{\footnote{Work supported by the Istituto Nazionale di Fisica Nucleare (INFN).}}

\def\Title#1{\begin{center} {\Large #1 } \end{center}}
\def\Author#1{\begin{center}{ \sc #1} \end{center}}
\def\Address#1{\begin{center}{ \it #1} \end{center}}

\newcommand\pubblock{\rightline{\begin{tabular}{l} \pubnumber\\
         \pubdate  \end{tabular}}}
\newenvironment{Abstract}{\begin{quotation}  }{\end{quotation}}
\newenvironment{Presented}{\begin{quotation} \begin{center} 
             PRESENTED AT\end{center}\bigskip 
      \begin{center}\begin{large}}{\end{large}\end{center} \end{quotation}}

\begin{document}
\begin{titlepage}
\pubblock

\vfill
\Title{Cosmic Ray Physics}
\vfill
\Author{ Domenico D'Urso\support}
\Address{\napoli}
\vfill
\begin{Abstract}
Cosmic ray story begins at the beginning of XX century. More then 100 years later, most of the main issues are still open questions, as sources, acceleration mechanism, propagation and composition.
There is a continuing fascination with the studies of cosmic radiation mostly from the several contradictions connected to its observation.
The radiation has an energy spectrum ranging from $\sim$ 1 GeV to beyond 10$^{20}$ eV with a flux going from 1 particle per m$^2$ per $\mu$s to less then 1 particle per km$^2$ per century, and so very different experimental techniques are needed to perform cosmic ray measurements in the different energy intervals.
In this contribution the actual experimental status of cosmic ray knowledge will be reviewed. 
\end{Abstract}
\vfill
\begin{Presented}
XXXIV Physics in Collision Symposium \\
Bloomington, Indiana,  September 16--20, 2014
\end{Presented}
\vfill
\end{titlepage}
\def\thefootnote{\fnsymbol{footnote}}
\setcounter{footnote}{0}

\section{Introduction}

The Earth's atmosphere is continuosly bombarded by extraterrestrial particles,
the so called Cosmic Rays ($CR$), which consist of ionized nuclei,
mainly protons, alpha particles and heavier nuclei. Most of them are
relativistic and a few particles have an ultrarelativistic kinetic energy,
extending  up to $10^{20}$ $eV$. 

$CR$ story starts at the beginning of the $20^{\mathrm{th}}$ century when
it was found that electroscopes
discharged even in the dark, well away from sources of natural
radioactivity. To solve the puzzle, in $1912$ Hess \cite{hesse} and
successively Kolh$\ddot{o}$rster \cite{kolhorster} made a series of manned
balloon flights, in which they measured the ionization of the atmosphere with
increasing altitude.

From $CR$ studies the elementary particle physics was born. Indeed, from
cosmic radiation track studies with cloud chamber,
Blackett and Occhialini \cite{blackett_occhialini} in $1933$ discovered
the positron and  in $1936$ Anderson and Neddermeyer
\cite{anderson_neddermeyer} announced the observation of particles
with mass intermediate between that one of the electron and the proton,
the muon.
With a new kind of instrument, nuclear emulsion, in $1947$ there was the
observation of the pion by Rochester and Butler \cite{rochester_butler} and so
 on till the $\Sigma$ particle discovery in $1953$ \cite{sigma_discovery}.

As testified from their history, $CR$ could be messenger of new physics 
and yet unknown particles. They are messengers of astrophysical
sites, even the most powerful ones. Via $CR$ studies it is possibile to infer
the properties of cosmic environment through which they propagate.
Furthermore, they are the only window on very high energy phenomena,  
well far from LHC energy scale, and  can be use to verify standard physical laws,
as the Lorentz invariance, in extreme conditions.

After more then 100 years of studies, most of the main related issues are still open questions, as sources, acceleration mechanism, propagation and composition.
These opened issues are a puzzle still today, whose solution involves astronomy
and cosmology, nuclear physics and elementary particle physics.

\section{Cosmic Ray Physics Case}

\begin{figure}[htb]
\centering
\includegraphics[height=2.5in]{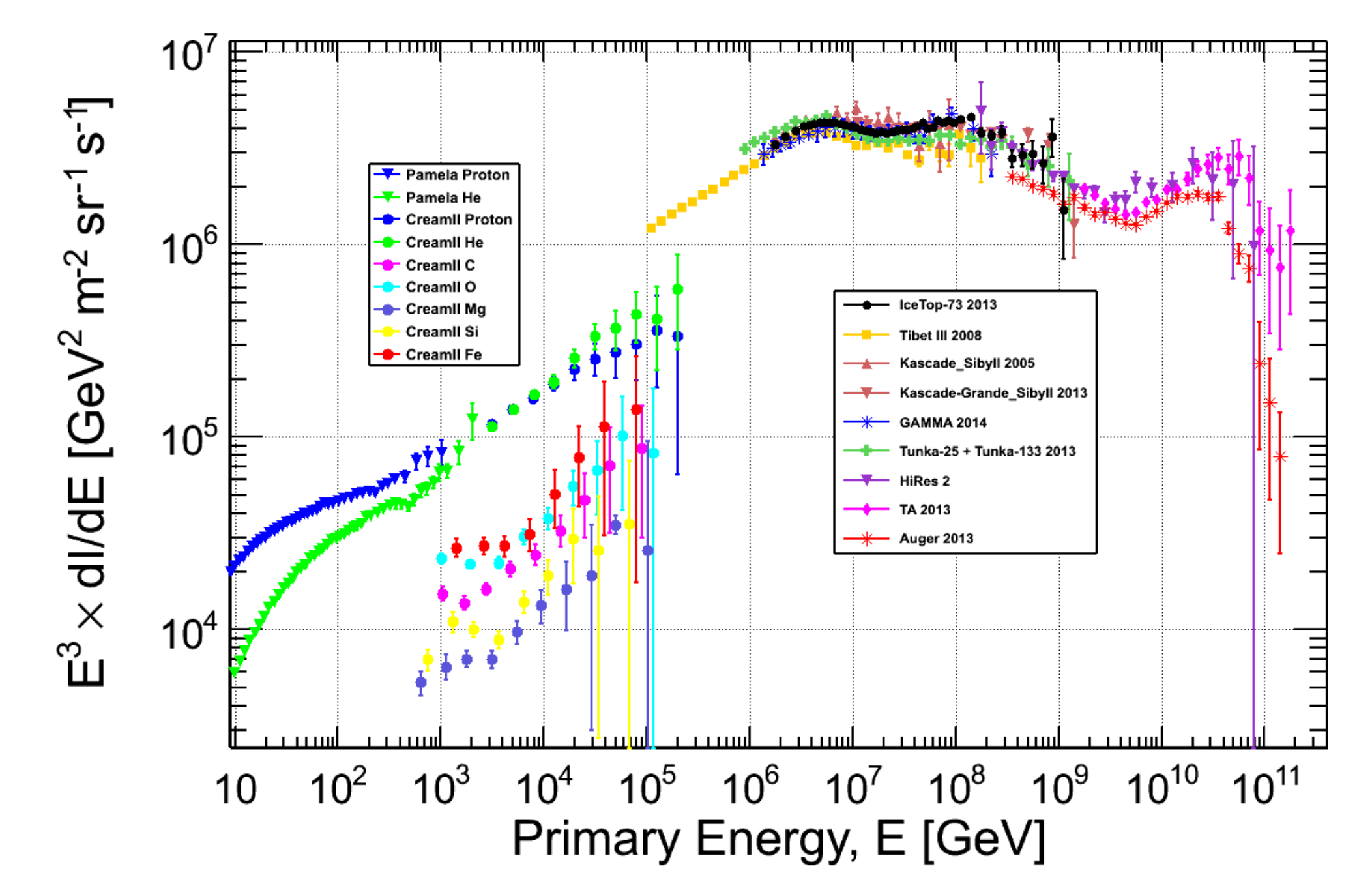}
\caption{All-particle energy spectrum from a compilation of measurements of the
 differential energy spectrum of $CR$ multiplied by E$^{3}$ to enhance spectrum 
 structures (plot extracted from \cite{all_particle_spectrum_plot}).}
\label{fig:all_particle_spectrum}
\end{figure}

The most striking feature of cosmic rays is their energy spectrum,
which spans a very wide range of energies with surprising regularity.
As it is possible to see in fig. \ref{fig:all_particle_spectrum}, the differential flux of
 the all-particle spectrum goes through 32 orders of magnitude along over
 12 energy decades. The regularity is broken only in few regions: the
\emph{knee} at about 3 $\times$ 10$^{15}$ eV, the \emph{second knee} at about 4 $\times$ 10$^{17}$, the \emph{ankle} at about 4 $\times$ 10$^{18}$ eV and the \emph{suppression} at the highest energies.
 Except a ``saturation'' region at low energies, cosmic ray spectrum can be well represented by power-law energy distribution

\begin{equation}
\frac{dN}{dE} \sim E^{-\gamma}
\end{equation}

where $\gamma$ $\sim$ $2.7$ up to the \emph{knee}, then $\sim$ $3.0$ up
to the \emph{second knee} where it becomes $\sim$ 3.3 up to the \emph{ankle}. Beyond the \emph{ankle}, the spectrum can be described with a $\gamma$ value of $\sim$ $2.6$ up to 
the \emph{suppression} region ($\sim$ 3 $\times$ 10$^{19}$).

Below the \emph{knee}, cosmic radiation is believed to originate within the galaxy and accelerated by supernovae explosion. 
As measured by direct experiments, $\sim$ 99.8\% of $CR$ are charged particles and only $\sim$ 0.2\% are photons and neutrinos.
99\% of charged particles are nuclei and 1\% are electrons and positrons. 
Among the nuclei, protons are $\sim$ 87\%, helium nuclei $\sim$ 12\% and heavier nuclei $\sim$ 1\%.
Relative abundances are quite similar to those measured from the Solar System.
The differences can be explained as result of spallation processes of primary particles propagating through the interstellar matter: excess of lithium, beryllium and boron nuclei;
excess of elements just less heavy then those of the iron group.  
In a standard scenario, where primary particle acceleration and propagation is due to magnetic fields (effects depend only on rigidity), the \emph{knee} is explained as the limit at which
galactic magnetic field can no more confine cosmic particles:
each radiation component is expected to have a \emph{knee} corresponding to the the rigidity at which it can be no more confined, from the lightest, protons and helium nuclei,
to the heaviest ones. 
In this scenario, the \emph{knee} is explained as the escaping limit of the light component of $CR$.
Recent experiments are pushing a new light on this region of the spectrum, with unprecedented precision measurements of particle spectra.
When it is not possible to perform direct measurements of primary particles
(typically for $E_{primary} > 10^{\mathrm{14}}$ $eV$),
the radiation composition is inferred from indirect observations. Then
conclusions are strongly dependent on models used to describe the data.
Different methods used for mass measurement usually give different answers.

As already anticipated, most of the main questions about $CR$ are still open: origin, composition and acceleration and propagation mechanisms.
Nowadays the $CR$ community is focusing on the following issues:
\begin{itemize}
\item particle abundances below the knee, searching for any hint of new physics;
\item connection between direct and indirect observations;
\item nature of spectrum features (knee, ankle,...);
\item galactic to extragalactic transition;
\item nature of the spectrum suppression at the highest energies.
\end{itemize}

Cosmic radiation of energy up to $10^{14}$ $eV$ could be studied directly
by detection of primary particle by means of balloon and satellite
experiments.
Going up with the energy, $CR$ flux becomes too low to use direct
detection, since it is impossible to employ wide area detectors
on balloons or in the space. Above $10^{14}$ $eV$ $CR$ have enough energy to initiate a cascade by interacting with the atmosphere,
whose products are detected at ground. At those energies, $CR$ 
are investigated
through the observation of these extensive air showers (EAS),
using particle detectors, at ground, of suitable area to measure
shower front and lateral distribution, the sampling
and the extension of the apparatus depend on the energy region one is interested in.

The direct detection allows to perform a particle identification and to have the energy scale
of the detector fixed by calibration at ground. The limit of measurement is the size of possible 
detectors and consequently the limited energy range. 
In the case of indirect measurements, it is possible to measure $CR$ up to 10$^{20}$ eV but 
all the measurements are based on modelling interactions of primary $CR$ with the atmosphere.

\section{Direct Measurements}

Direct measurements are performed by stratospheric balloons or space experiments.
In the last decades, direct experiments on cosmic rays received a push 
forward by the possibility of conducting experiments on board of long duration balloon flights, satellites and on the International Space Station. 
The increase in the collected statistics and the technical improvements
 in the construction of the detectors allows to measure the fluxes 
at higher energies with a reduced discrepancy among 
 different experiments respect to the past.
Due to the statistics of particle flux, the sensitivity of different experiments are 
determined by the combination of detector livetime and acceptance.
Balloon experiments usually have a larger acceptance but a shorter livetime, while
space experiments can have a larger livetime but a reduced detector size due to
size/weight limits imposed by space mission constraints.
In the last years there were no updates from balloon experiments. 
For a review of direct measurements of $CR$ by balloon experiments one can 
refer to \cite{direct_measurements_review}.

Recently important  observations have been reported about the proton and helium spectrum by PAMELA \cite{pamela_review}
and by AMS-02 \cite{ams_general_reference} about the electron and positron spectra 
In the following the two detectors will be described and their main results discussed.

\subsection{PAMELA and AMS-02 experiments and results}
\begin{figure}[htb]
\centering
\begin{minipage}{0.4\linewidth}
  \includegraphics[height=2.2in]{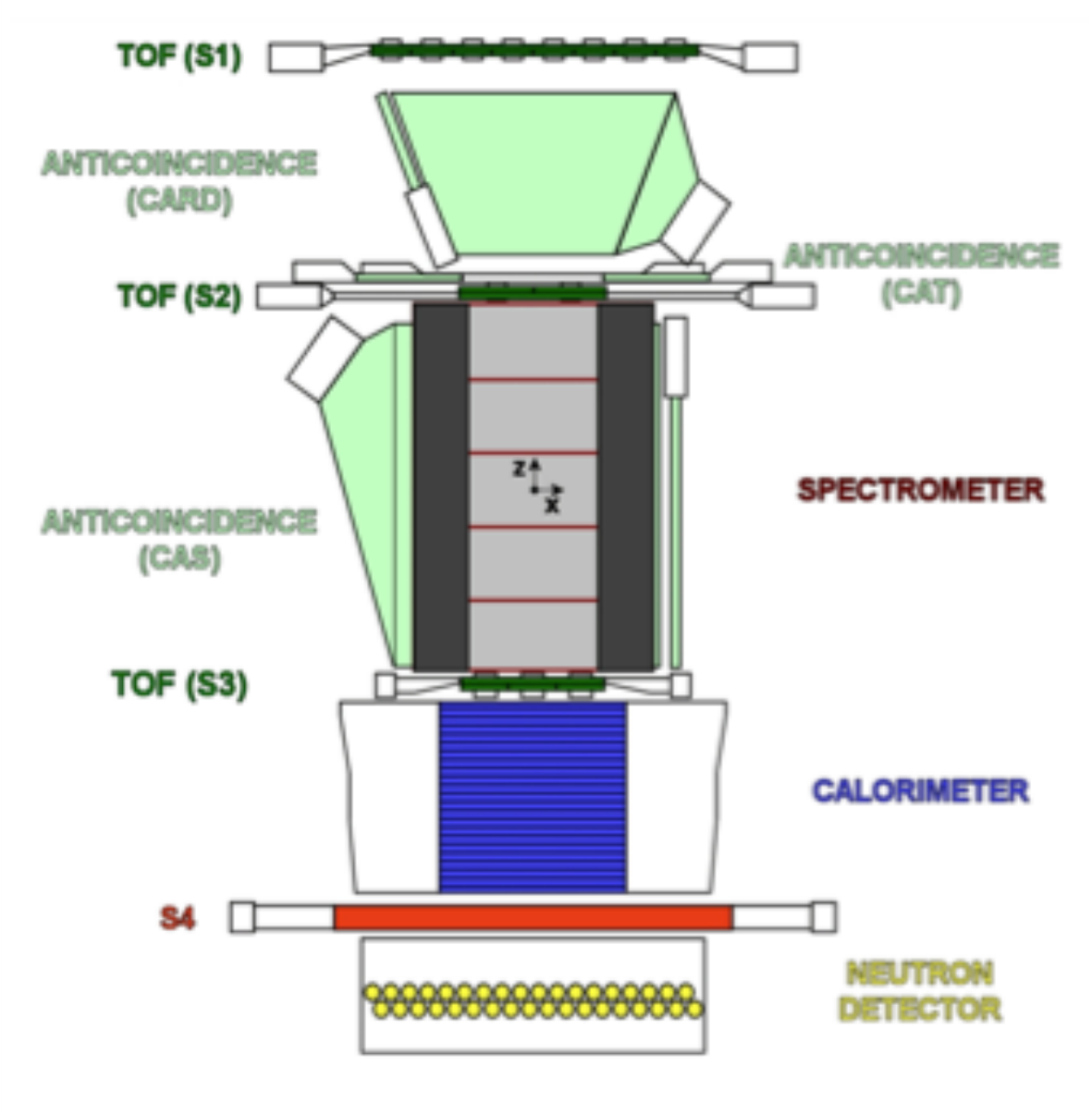}
  \caption{Schematic view of the PAMELA apparatus.}
  \label{fig:pamela_scheme}
\end{minipage}
\begin{minipage}{0.4\linewidth}
  \includegraphics[height=2.2in]{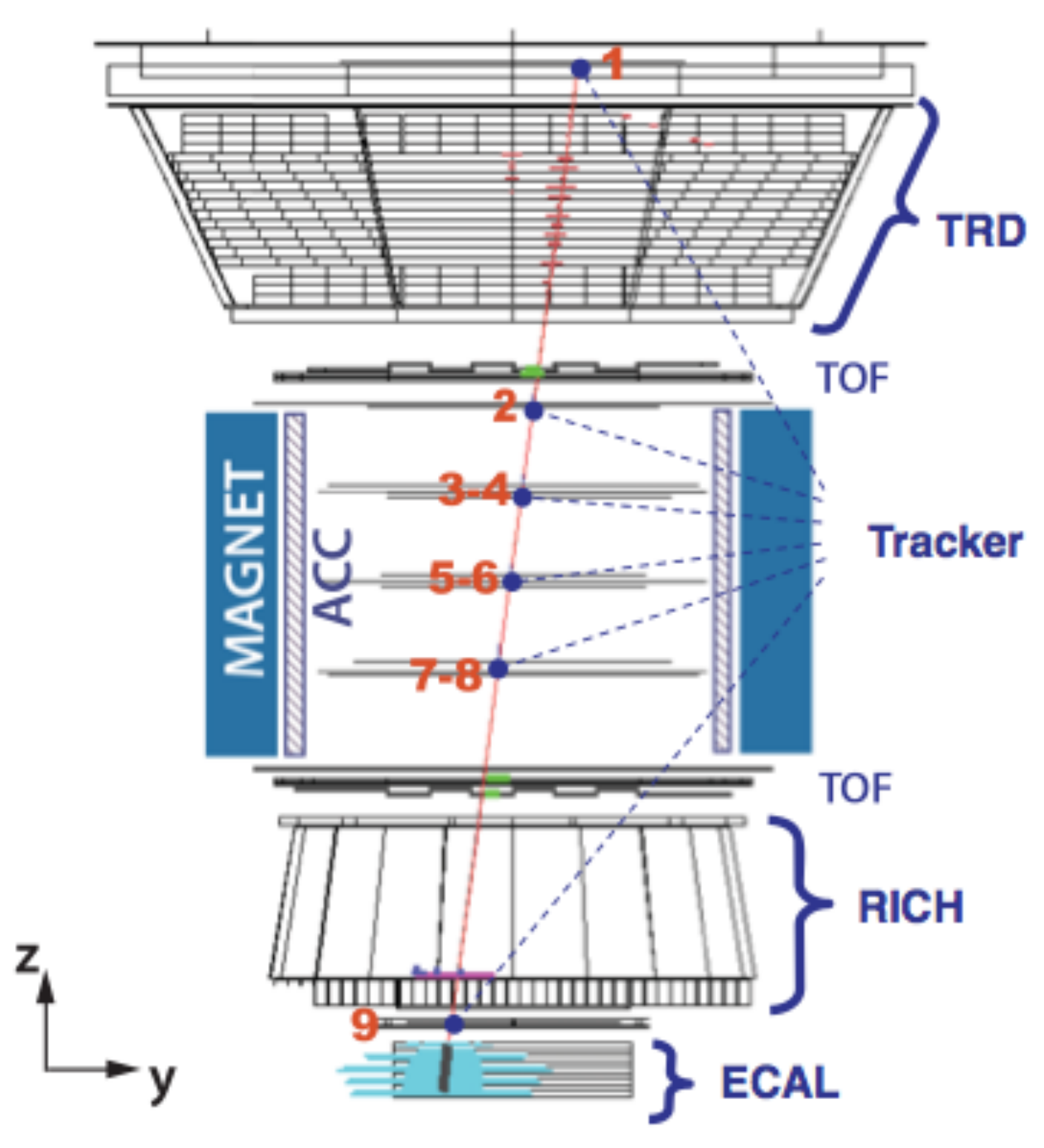}
  \caption{Schematic view of the AMS-02 apparatus traversed by a 1.03 TeV electron event.}
  \label{fig:ams_scheme}
\end{minipage}
\end{figure}

PAMELA (Payload for Antimatter Matter Exploration and Light-nuclei Astrophysics) is a satellite-borne experiment designed to study charged particles in the cosmic radiation.
A schematic view of the apparatus is shown in fig. \ref{fig:pamela_scheme}. The central part of the apparatus is a magnetic spectrometer, consisting of a permanent magnet 
and a 6 double layer of silicon detector planes to measure the rigidity and the ionisation energy losses, $dE/dx$.  
Below the spectrometer an electromagnetic calorimeter (ECAL) is installed to measure energy particle and to perform particle identification . 
The apparatus is equipped with a Time of Flight (ToF)  system of plastic scintillators for the trigger, with a capability of 12 independent measurements of particle velocity, $\beta$ = $\nu$/c.
The ToF allows to discriminates between down-going and up-going particle, enabling the spectrometer to determine the sign of the particle charge, and 6 independent 
$de/dx$ measurements.
There is also an anticoincidence system to identify in the offline analysis apse triggers and multi-particle events corresponding to secondary particles produced in the detector.
In the bottom part of the apparatus, there is a neutron detector(ND) that detects the neutron production associate with hadrons to increase the electron-proton discrimination.

The AMS-02 (Alpha Magnetic Spectrometer) is a general purpose high energy particle physics detector installed on the Internation Space Station.
The layout of AMS-02 detector is shown in fig \ref{fig:ams_scheme}. 
It consists of nine planes of precision silicon tracker, a transition radiation detector (TRD), four planes of time of flight counters (TOF), a permanent magnet, an array of anticoincidence counters (ACC), placed around the inner tracker, a ring imaging $\breve{C}$erenkov detector (RICH) and an electromagnetic calorimeter (ECAL) of 17 radiation lengths. The figure also reports a high-energy electron of 1.03 TeV detected by AMS.
The tracker accurately determines the trajectory and absolute charge (Z) of particles. 
The TRD is designed to use transition radiation to distinguish between electrons and protons, and $dE/dx$ to independently identify nuclei.
The coincidence of signals from all of the four ToF planes provides a charged particle trigger for the apparatus.
The ACC system around the magnet and the inner tracker allows to reject events corresponding to particles entering or leaving the inner tracker volume transversely.
The RICH is designed to have an independent measurement of  the magnitude of the charge of cosmic rays and their velocities with a precision of $\Delta \beta$/$\beta$  $\sim$ 1/1000.
The ECAL allows to measure particle energy and it provides an independent discrimination between electrons and protons.

\begin{figure}[htbp]
\centering
\includegraphics[width=0.3\textwidth]{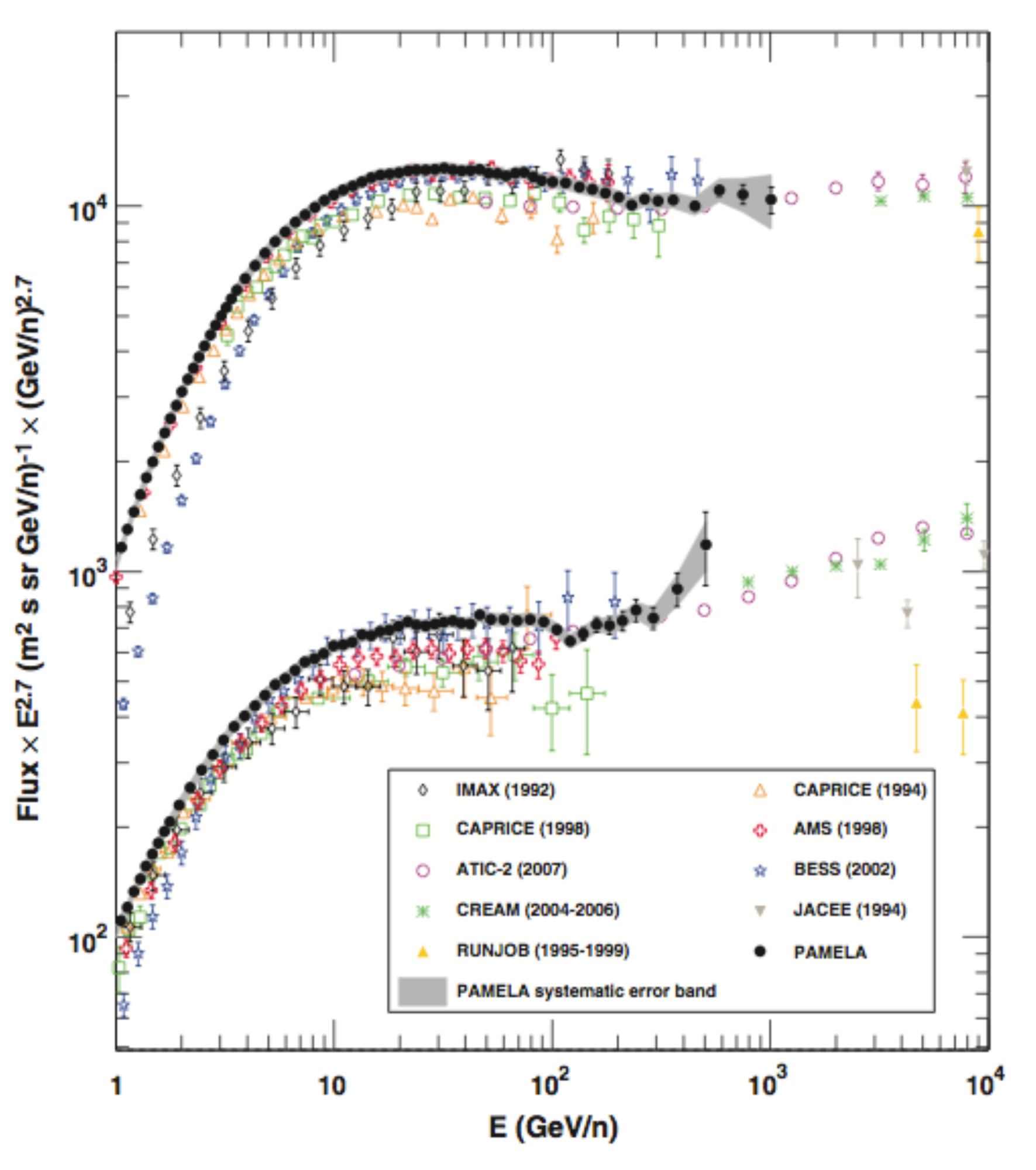}
\includegraphics[width=0.5\textwidth]{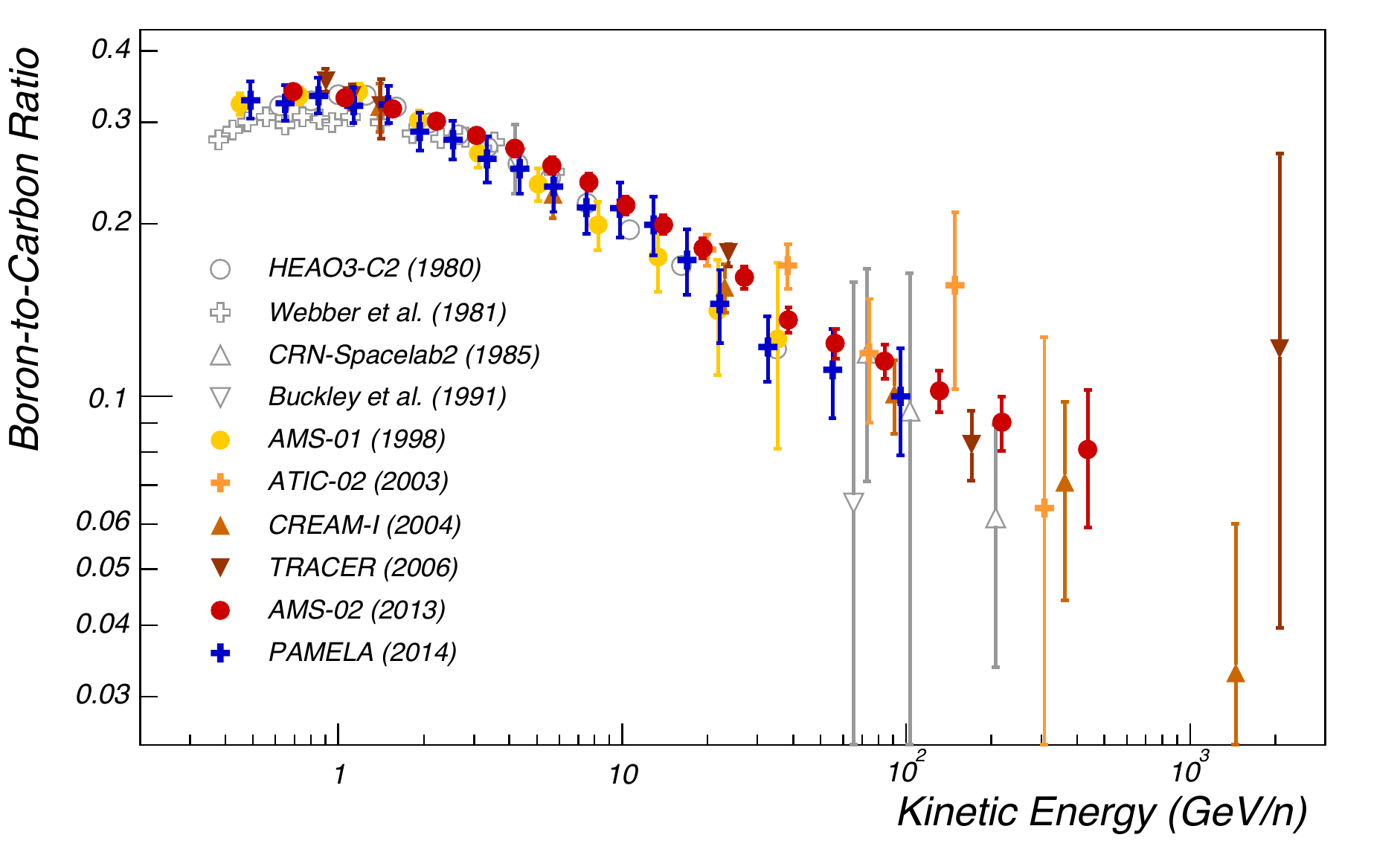}
\caption{Proton and helium absolute fluxes (left plot) measured by PAMELA  \cite{phe_pamela} above 1 GeV per nucleon 
and boron-to carbon flux ratio (right plot) measured by PAMELA  \cite{b_c_pamela}  and AMS-02 \cite{b_c_ams}.  
Results from previous measurements \cite{phe_old} \cite{b_c_previous} are also shown.}
\label{fig:phe_pamela}
\end{figure}

Protons and helium nuclei are the most abundant $CR$ species and hence they allow to understand the detector limits in terms of tracker resolution and systematic uncertainties.
Recently a measurements of proton and helium spectra has been published by PAMELA \cite{phe_pamela}.
In the left plot of fig. \ref{fig:phe_pamela} proton and helium fluxes are shown multiplied by a power of the energy, E$^{2.7}$, to enhance spectral features, compared with a few of previous measurements. 
In this energy region, in a scenario based on a shock diffusion acceleration model and a diffusive propagation in the Galaxy \cite{standard_acc_prop_model}, 
the spectrum is expected to have a featureless behaviour and to be well fitted by a single power law with similar spectral indices for protons and heavier nuclei. 
For energy greater than 30 GV (above the influence of solar modulation), PAMELA  observed a different spectral shape for proton and helium spectra. Furthermore data can not be described by a single power law model, a change in the spectral index has been observed around 230 GV (the single power law hypothesis is rejected at the 95 \% confidence level).
The measured spectral indices are $\gamma_1$p = 2.85 $\pm$ 0.015(stat) $\pm$ 0.004(sys) and $\gamma_1$He = 2.766 $\pm$ 0.01(stat) $\pm$ 0.027(sys) below 230 GV for protons and helium nuclei respectively, while they become  $\gamma_2$p = 2.67 $\pm$ 0.03(stat) $\pm$ 0.05(sys) and $\gamma_2$He = 2.48 $\pm$ 0.06(stat) $\pm$ 0.03(sys)  above 230 GV.
Those observations challenge  the ``standard'' scenario: models do not reproduce data across the full-rigidity region 
and they do not predict a significant difference between proton and helium indices. 
New features could be interpreted as an indication of different population of $CR$ sources \cite{pamela_explanation_source} or of a new physical phenomenon in the propagation \cite{pamela_explanation_propagation} . 


The relative abundances of $CR$ species provide information about cosmic-ray transport within the Galaxy: 
$CR$  such as carbon and oxygen may interact with the interstellar medium to produce secondary fragments such as lithium, beryllium and boron. 
The ratio of secondaries to primaries $CR$ can be used to estimate the amount of traversed interstellar matter.
One of the most sensitive quantities is the ratio of boron to carbon, because boron is purely secondary and its main progenitors, carbon and oxygen, are primaries. 
The shape of this ratio is highly sensitive to propagation coefficients.
Due to their similar charge, the B/C ratio is the less affected by systematics or solar modulation.
Recently PAMELA and AMS-02 reported a measurement of B/C ratio flux \cite{b_c_pamela}  \cite{b_c_ams}. 
In the right plot of fig. \ref{fig:phe_pamela} the B/C ratio as a function of kinetic energy per nucleon measured by PAMELA and
AMS-02 is shown with data of previous experiments. New data fix tighter constraints for propagation models.
Anyway, at high energy, the main limitation for the ratio measurement is the statistics but AMS has collected only 10\% of the expected statistics. 
The B/C behaviour at high energy will be become more clear with more data.

Electrons and positrons are only $\sim$ 1\% of cosmic radiation but they provide important informations regarding the origin and propagation
of $CR$. 
In the last months, AMS-02 published the measurements of positron and electron fluxes \cite{eflux_ams} and an update of positron fraction \cite{fraction_ams}, 
fixing the limit of present knowledge of electron/positron radiation component. 

\begin{figure}[htbp]
\centering
\includegraphics[width=0.41\textwidth]{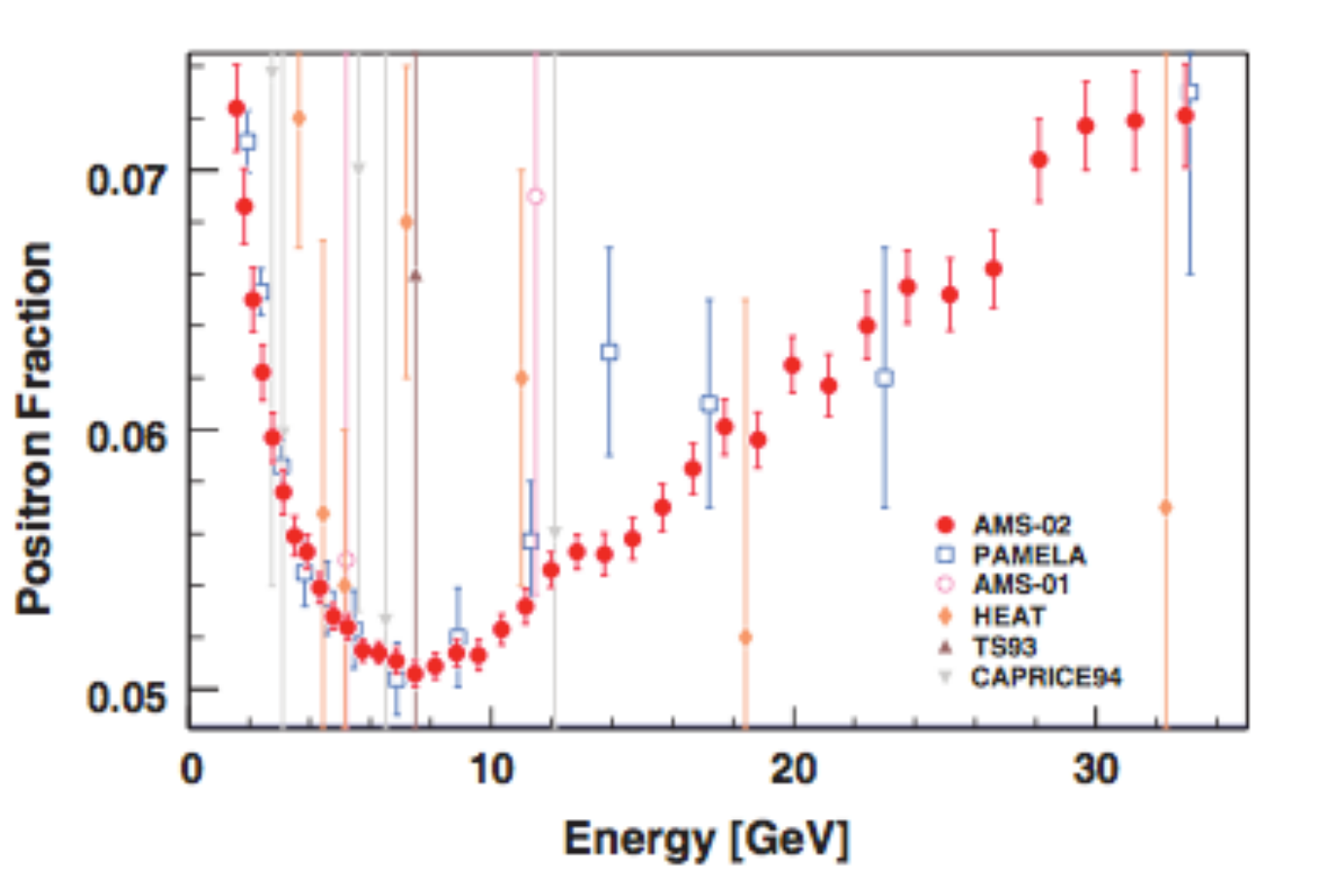}
\includegraphics[width=0.45\textwidth]{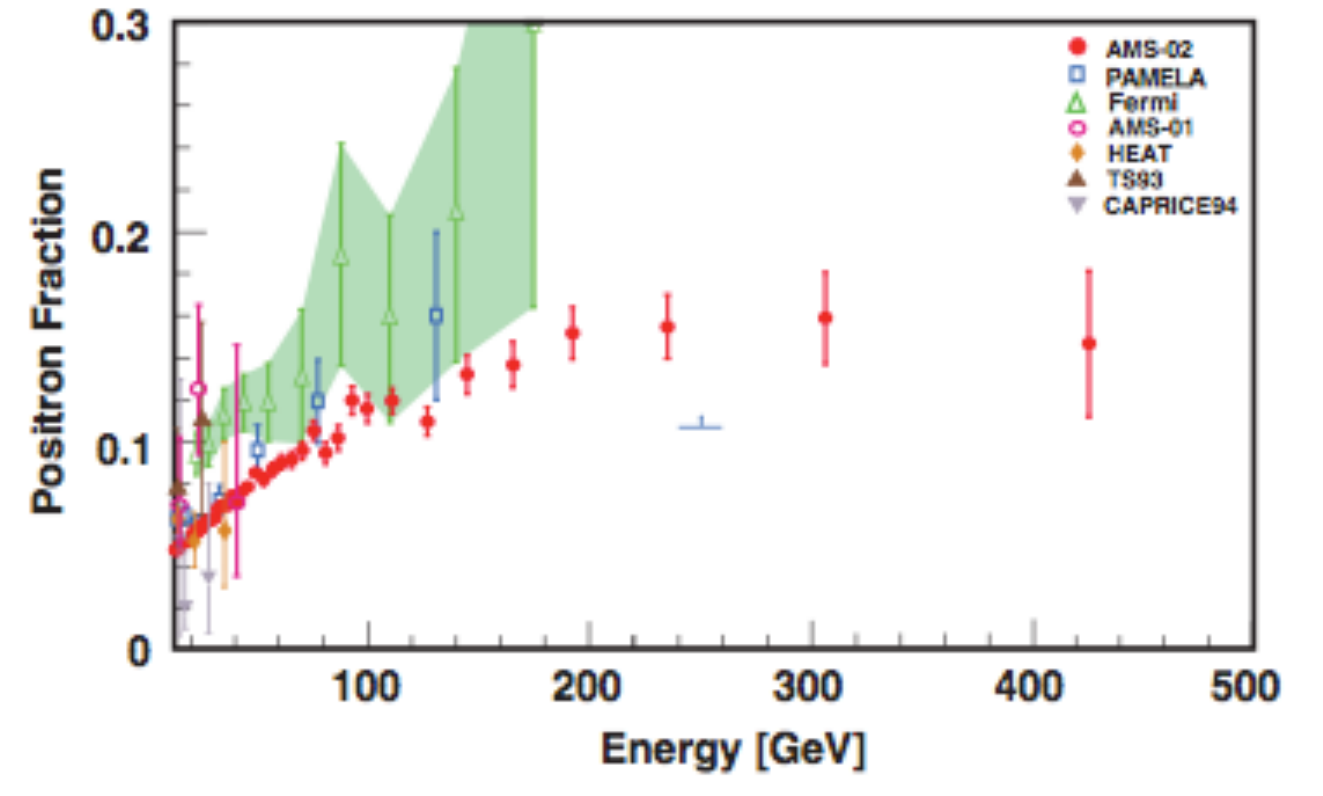}
\caption{AMS-02 positron fraction from 1 to 35 GeV (left) and from 10 to 500 GeV, compared with previous measurements.} \label{fig:e_ams_fraction}
\includegraphics[width=0.3\textwidth]{e_flux_ams_1}
\includegraphics[width=0.3\textwidth]{e_flux_ams_2}
\caption{ Global (left plot) of AMS-02 (a) electron, from 0.5 to 700 GeV, and (b) positron fluxes, from 0.5 to 500 GeV, multiplied by E$^{-3}$.
In the right plot a detailed view from 0.5 to 200 GeV is shown.
Also shown are data from previous experiments.}
\label{fig:e_flux_ams}
\end{figure}

The positron fraction is defined as the ratio of the positron flux to the combined flux of positrons and electrons.
In fig. \ref{fig:e_ams_fraction} the measured positron fraction by AMS-02 is presented as a function of the energy, compared with previous measurements, from 1 to 35 GeV (left plot)
and from 10 to 500 GeV (right plot).
As expected from diffuse positron production, there is a rapid decrease from 1-8 GeV.
From 10 to ~200 GeV a steadily increasing has been confirmed and above 200 GV the fraction seems to flatten out.
An upper limit on dipole anisotropy amplitude d$\leq$0.030@95\%CL (E$>$16 GeV) has been also reported.
In  this scenario, the measurement of the  \emph{end} of positron fraction will be crucial to understand the nature of the positron excess.

In fig. \ref{fig:e_flux_ams} are shown the separated fluxes for electrons and positrons measured by AMS-02, from 0.5 to 700 GeV
and from 0.5 to 500 GeV respectively.
Neither the electron flux nor the positron flux can be described by single power law over the entire range.
Above $\sim$ 20 GeV and up to 200 GeV the electron flux decreases more rapidly with energy than the positron flux.  
This is not consistent with only the secondary production of positrons \cite{ep_secondary}.
So the rise in the positron fraction seems to be due 
to the hardening of the positron spectrum and not to the softening of the electron spectrum.

\section{Indirect Measurements}
The indirect measurements of $CR$ are based on the observation of $EAS$ observables:
particles at ground, like electrons, mouns and hadrons;
$\breve{C}$erenkov light, used in the energy range 10$^{14}$-10$^{16}$ eV;
fluorescence light, used for energies higher than 10$^{17}$ eV.
Recently has been proven also the possibility of using radio signals.   
The main issue of $CR$ detected by means of indirect techniques is the understanding of 
the EAS generated by a primary particle whose energy is estimated to exceed 10$^{18}$ eV (Ultra High 
Energy Cosmic Rays, UHECR).
In the last years, new results have been published about the energy spectrum, the mass composition and the 
distribution of arrival directions  of primary particles by the Auger \cite{auger_exp} and the Telescope Array (TA) \cite{ta_exp} experiments.
In the following Auger and TA experiments will be described and their results will be discussed. 
For a history and a complete review of $CR$ indirect detection techniques one can refer to \cite{nagano_watson, watson_kampert}.

\subsection{Auger and TA experiments and results}
The two experiments are the last UHECR generation experiments, based on the use of an \emph{hybrid} detection technique:
a combined use of a surface detector (SD)  and of a nitrogen fluorescence detector (FD).
Employing two complementary techniques to observe EAS, SD and FD are able to perform independent measurements on the same shower: 
the lateral and temporal distribution of shower particles at the ground level with SD, 
the air shower development in the atmosphere (the longitudinal profile) above the surface array with FD.
Both techniques have limits and advantages.
FD allows to have a nearly calorimetric energy measurement, with a direct view of shower evolution, but it has only 15\%
of duty cycle and its acceptance grows up with energy and depends on atmosphere conditions.
SD has a duty cycle of 100\% and an acceptance flat above threshold, but its estimation of primary energy is based on Monte Carlo studies.   
 
The use of two detection techniques allows to have: 
an inter-calibration of the two methods, to study their systematic uncertainties;
an energy spectrum estimation almost model independent;
an enhanced composition sensitivity, FD directly measures the depth at which the shower reaches its maximum (X$_{max}$) while
SD measures particle densities and time width of the shower front at ground level, all fingerprints of primary's nature.

The Pierre Auger Observatory is located in the Pampa Amarilla upland. 
The SD of the Pierre Auger Observatory \cite{auger_sd} consists of 1600 water Cherenkov tanks on a regular hexagonal grid 
with a distance of 1500 m between tanks, covering a total area of 3000 km$^2$. 
At an altitude of 1400 m above sea level, corresponding to a vertical atmospheric depth of 875 g cm$^{-2}$,
The area is overlooked by 4 fluorescence detectors (eyes) \cite{auger_fd}, disposed at the edges of the surface array.

TA is located in the high desert in Millard County, Utah.
It consists of a SD of more then 500 scintillator detectors disposed on a 1.2 km square grid,  covering 700 km$^2$. 
The area is overlooked by 3 telescope stations.

\begin{figure}[htbp]
\centering
\includegraphics[width=0.43\textwidth]{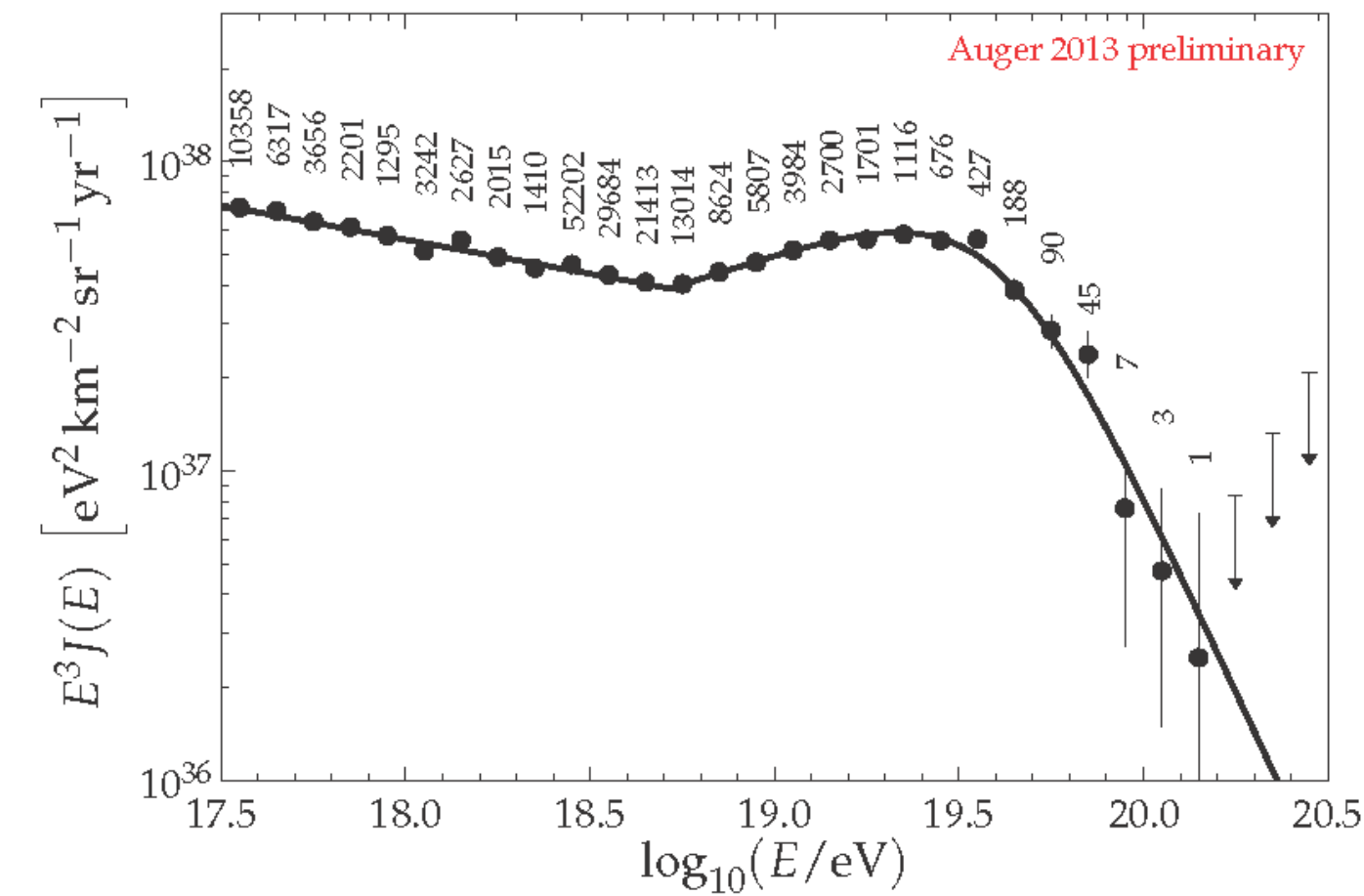}
\includegraphics[width=0.35\textwidth]{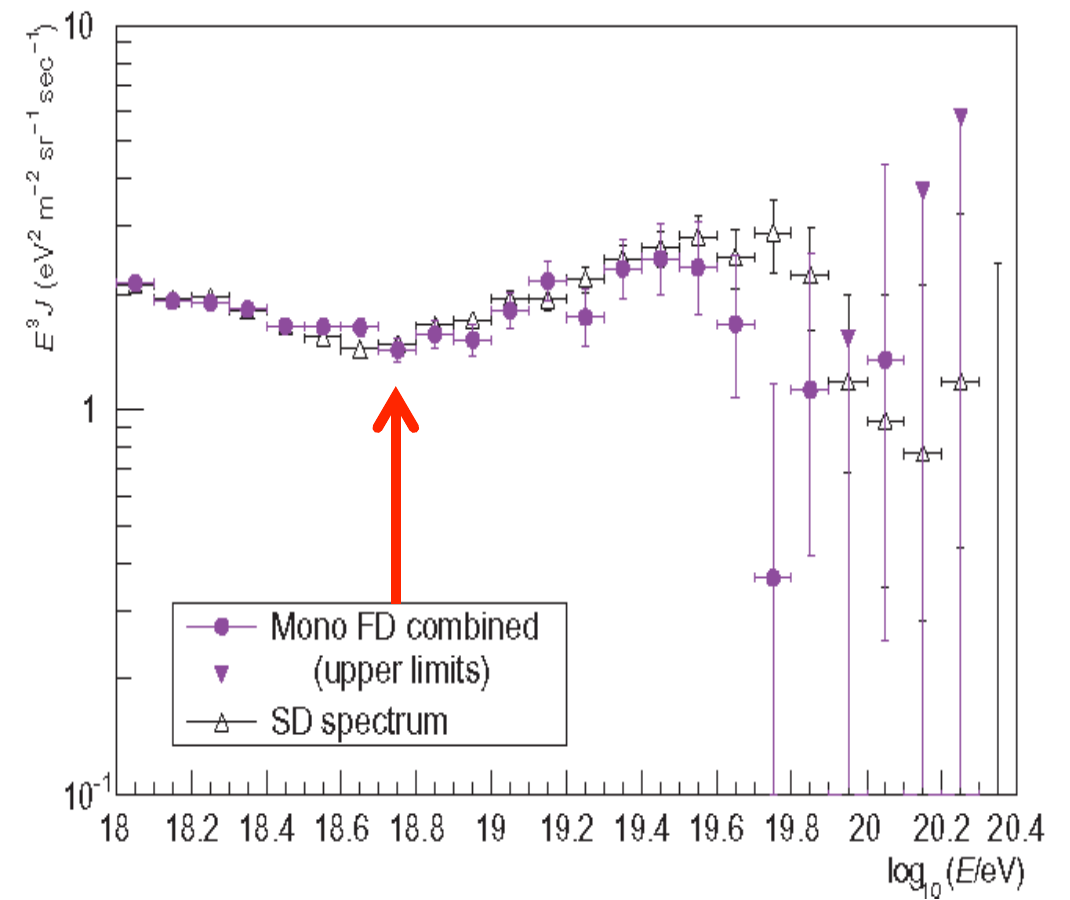}
\caption{Energy spectrum as measured at the Pierre Auger Observatory (left) and by the Telescope Array experiment (right) using the Mono FD combined and the SD measurements. 
In the case of Auger, numbers give the total number of events inside each bin and  the last three arrows represent upper limits at 84\% CL, for TA upper limits are given at 68\% CL.}
\label{fig:auger_ta_spectrum}
\end{figure}
Recents measurements of energy spectrum have been reported by the Pierre Auger observatory at the ICRC 2013 \cite{spectrum_auger}
and by the TA experiment \cite{spectrum_ta} (see fig. \ref{fig:auger_ta_spectrum}).
The Auger Collaboration presented an update of his spectrum measurement, combining the hybrid spectrum (hybrids events) with the SD 
spectrum (SD events whose energy has been calibrated with FD), covering the range from 10$^{17.5}$ to 10$^{20.5}$ eV. 
To characterize  the spectral features data have been described with a simple power law below the ankle, 5$\times$ 10$^{18}$ eV
 and a power law with smooth suppression above. The obtained spectral index above the ankle is 3.23 $\pm$ 0.01(stat) $\pm$ 0.07 (sys)
 and 2.63 $\pm$ 0.02(stat) $\pm$ 0.04 (sys) above. The suppression is seen with a significance of 20$\sigma$.
The TA Collaboration presented a measurement FADC-based FDs in monocular mode, from 10$^{18}$ to 10$^{20.2}$ eV.
TA identify  the ankle at 10$^{18.75}$ eV and a spectrum suppression at 10$^{19.50}$ eV with a significance of 3.2 $\sigma$. 
Experiments have a difference in their energy scale of about 14\%, compatible with their systematic (14\% for Auger and  21\% for TA). 
Both confirm the spectrum structures of ankle and the suppression, however their origin is yet to be determined.
These features can originate from the interaction of primaries with the CMB: 
attenuation due to pair production can explain the ankle and pion production from protons-CMB interactions or photo-disintegration of nuclei can produce the suppression.
Alternatively, the structures derive from the spatial distribution of sources and their acceleration 
mechanisms. 
In that case the ankle could identify the transition point from a galactic dominated radiation to an extra-galactic dominated one.
Several scenarios have been put forward to explain the spectrum (for an overview see \cite{explanation_review}).

\begin{figure}[htbp]
\centering
\includegraphics[width=0.40\textwidth]{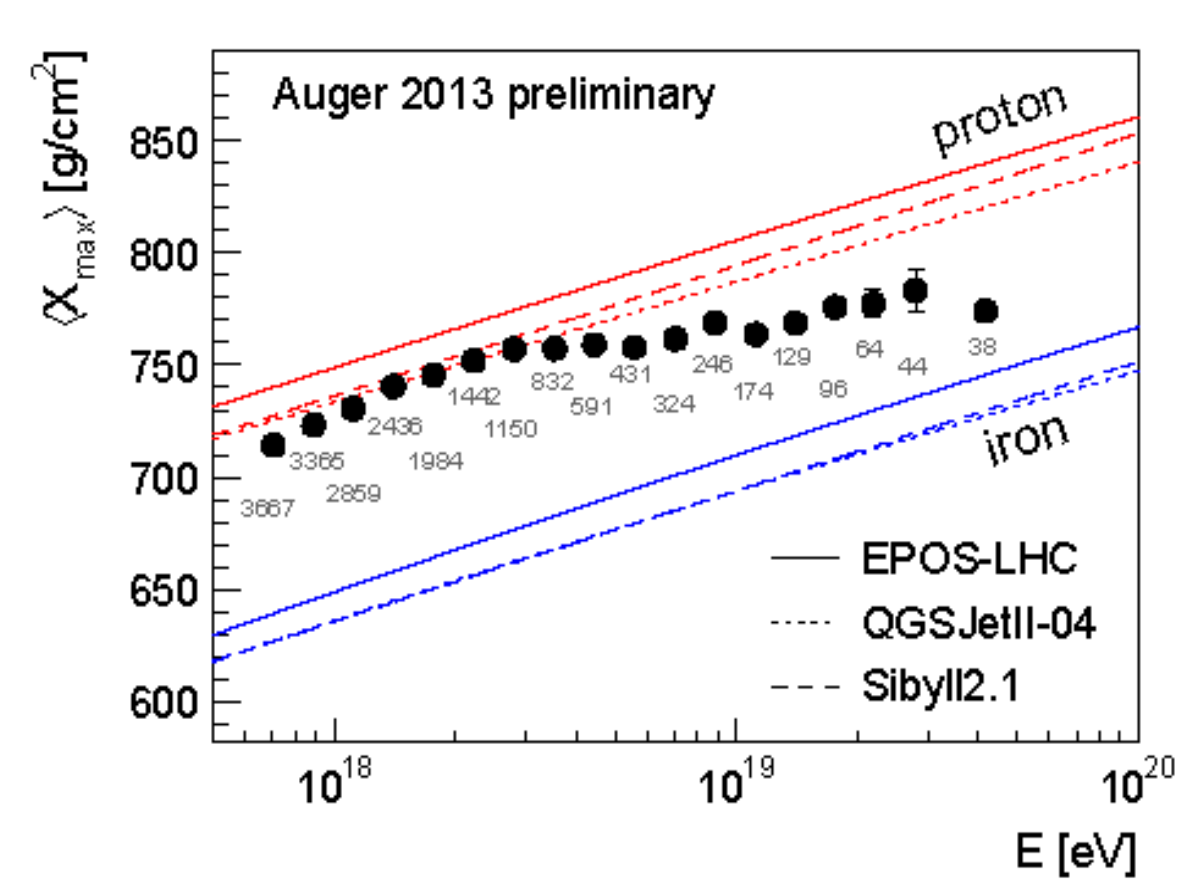}
\includegraphics[width=0.36\textwidth]{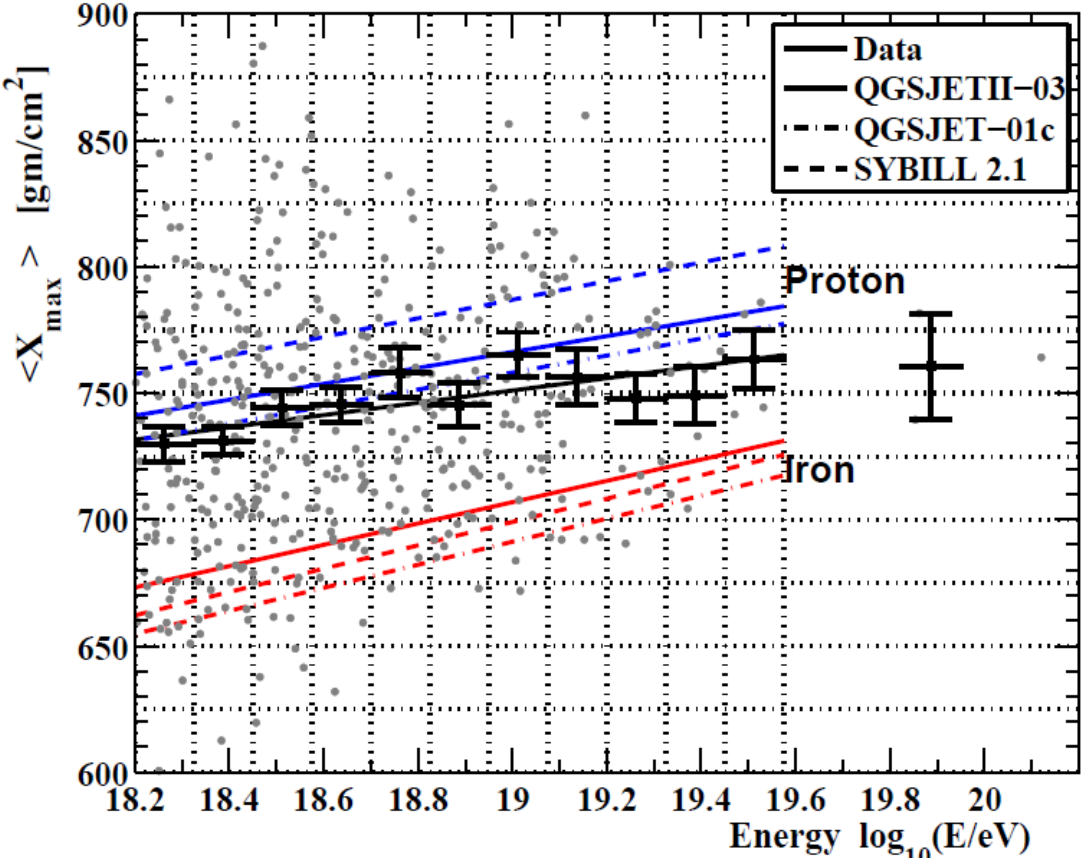}
\caption{Evolution of $\leftarrow	 X_{max} \rightarrow$ as a function of the energy measured by the Pierre Auger Observatory (left)
and the Telescope Array experiment(right) compared with simulation expectations.}
\label{fig:auger_comp}
\end{figure}

Important part of this puzzle is the mass composition measured in this energy range.
The Pierre Auger Collaboration reported \cite{auger_composition} an update of the evolution with energy of the first two moments 
of the X$_{max}$ distributions.
As one can see from left plot of fig. \ref{fig:auger_comp},   
Data clearly indicate a change of behaviour at a few EeV, i.e. in the Ankle region.  
Under the hypothesis that no new interaction phenomena in the air shower development come into play, 
the data clearly support the hypothesis of a mass composition evolving in the Ankle region.
In the case of TA data, see right plot in fig. \ref{fig:auger_comp}, 
a light, nearly protonic, composition is in good agreement with the data, in the whole energy range.
Detailed comparison of the two measurements and of the two analysis techniques to understand 
the mutual differences \cite{auger_ta_comparison} are in progress.

The observation of structures in the arrival direction distribution of primaries is also related to the mass of the radiation, it affects
the angular scale at which a possibile correlation could be observed, and directly to the origin of the radiation (Are there nearby sources?).
Searches of possible anisotropies have been done by both hybrid experiments.
In particular, the Pierre Auger Collaboration claimed a possible correlation between the distribution of arrival directions and an AGN catalogue \cite{auger_correlation}.
Auger Collaboration found evidence of anisotropy in the arrival directions of cosmic rays above the Greisen-Zatsepin-Kuzmin energy threshold, 
6 $\times$ 10$^{19}$ eV. 
The anisotropy was measured by the fraction of arrival directions that are less than
3.1$^{\circ}$ from the position of an active galactic nucleus within 75 Mpc (using the V$\acute{e}$ron-Cetty and V$\acute{e}$ron 12th catalog).
In their last publication on the subject \cite{auger_anisotropy}, with 69 events with energy greater then 55 EeV, they reported a degree of correlation of 38$^{+7}_{-6}$\%, against a 21\% expected in the hypothesis of an isotropic flux.
Furthermore, an excess around Cen A has been observed, 18.8 \% of events lie within 18$^{\circ}$ from the source, while 4.7\% is expected from an isotropic flux.
A cluster of arrival direction has been observed also by the TA experiment \cite{ta_anisotropy}.
Using $CR$ with energy greater than 57 EeV (87 events), they found a cluster centred at (R.A.=146.7$^{\circ}$, decl.=43.2$^{\circ}$), with a statistical significance of 5.1 $\sigma$. 
They also estimated the probability of such a hotspot to appear by chance in an isotropic sky to be 3.7 $\times$ 10$^{-4}$.


\section{Discussion and conclusions}
After more then 100 years of studies and experiments, the understanding of cosmic radiation is still a puzzle to be solved.

In the last decades, direct experiments on cosmic rays received a push 
forward by the possibility of using long duration balloon flights, satellites and the International Space Station. 
The increase in the collected statistics and the technical improvements of apparatus of the most recent experiments
offered the opportunity to have very important results on the cosmic radiation, especially from space-borne experiments.
Accurate measurements of positron fraction and of electron and positron fluxes have been performed.
A clear positron excess has been established from 10 to 200 GeV. AMS-02 Collaboration has observed  a flatting above 200 GeV
and only the determination of the \emph{end} of the excess will help to understand the nature of the excess and if is an indirect sign of dark matter.
Detailed measurements of proton and helium spectra from PAMELA Collaboration show several unexpected features:
different spectral indices for the primaries and a change of the spectral index around 230 GeV for both. 
Only a more accurate measurement will clarify the existence of those structures (AMS-02 results are expected by the beginning of 2015).
A very accurate measurement of boron-to-carbon ratio has been performed by the AMS-02 and PAMELA Collaborations, putting tighter constraints 
on propagation models of $CR$ in the Galaxy. 
While PAMELA experiment is over, AMS-02 will take data till the decommissioning of the International Space Station.
In the present AMS-02 publications only 10\% of the total available statistics is used, 
hence mayor contributions are expected from AMS-02 in the future.
Furthermore, there are several new space based experiments that will be installed on the ISS, that will become a kind of $CR$ Observatory. 
Those experiments will extend the energy range of direct measurements of galactic $CR$, leading to the connection between direct and indirect observations.

At higher energies, a full explanation of all spectral features is still missing. 
The last generation of UHECR experiments, the Auger Observatory and Telescope Array experiment, confirm the \emph{end} of the spectrum
but its nature is unclear. There is still to be understood if there is a transition from a galactic dominated radiation to an extra-galactic dominated one
and if it occurs at which energy and in which way.
Spectral features could originate from the interaction of primaries with the CMB, then suppression would be explained in a GZK-scenario.
Or they could derive from the distribution of sources and their acceleration mechanisms and in that case the suppression would be explained in a source-scenario. 
UHECR mass composition is still an open issue and data reported from Auger and TA Collaborations are not in agreement. 
For the future, there are many upgrades planned for Telescope Array experiment and the Auger Observatory to understand the 
origin of the flux suppression, to identify the proton contribution at the highest energies and to improve comprehension of EAS development.


\end{document}